\documentstyle[aps,epsf]{revtex}
\begin{document}
\title{The effect of finite-range interactions in classical transport theory}
\author{Sen Cheng and Scott Pratt\\
Department of Physics and National Superconducting Cyclotron 
Laboratory\\
Michigan State University\\
East Lansing Michigan, 48824}
\author{Peter Csizmadia\\
RMKI Research Institute for Particle and Nuclear Physics\\
P.O. Box 49, Budapest, 1525, Hungary}
\author{Yasushi Nara\\
Riken BNL Research Center, Brookhaven National Laboratory, Upton, NY 11973}
\author{D\'enes Moln\'ar and Miklos Gyulassy\\
Department of Physics, Columbia University\\
538 120th St., New York, NY 10027}
\author{Stephen E. Vance\\
Physics Department, Brookhaven National Laboratory, Upton, NY 11973}
\author{Bin Zhang\\
Department of Chemistry and Physics, Arkansas State University\\
P.O. Box 419, State University, Arkansas 72467-0419}
\date{\today}


\maketitle

{\abstract The effect of scattering with non-zero impact parameters between
consituents in relativistic heavy ion collisions is investigated. In solving
the relativistic Boltzmann equation, the characteristic range of the collision
kernel is varied from approximately one fm to zero while leaving the mean-free
path unchanged. Modifying this range is shown to significantly affect spectra
and flow observables. The finite range is shown to provide effective
viscosities, shear, bulk viscosity and heat conductivity, with the viscous
coefficients being proportional to the square of the interaction range.  }

\section{Introduction}

A principal goal of relativistic heavy ion collisions is to experimentally
discern bulk properties of the excited vacuum. To accomplish this aim it is
imperative that one understands the implications of the finite size and
lifetime of the global reaction. Two microscopic length scales govern the
importance of finite-size effects, the mean free path and the range of
interaction. Microscopic models, e.g., those based on the Boltzmann equation,
easily incorporate the effects of a finite mean free path. Such effects can be
linked to viscous terms in analogous hydrodynamic descriptions.

In intermediate-energy collisions, where excitation energies are tens of MeV
per nucleon, the role of the finite range to the strong interaction has been
studied in its relation to the surface energy of nuclear matter. In such
Boltzmann descriptions the binding energy of nuclear matter is introduced via
the mean field, with the coarseness of the mean field mesh being adjusted so
that the effect of the interaction range is effectively tuned to reproduce the
surface energy of nuclear matter. In non-relativistic molecular dynamics, the
effects of hard-sphere interactions have also been investigated. In this case
the size of the spheres represents a length scale which can strongly affect
bulk properties of the matter at high density. However, in the context of a
Boltzmann description, where $n$-body correlations are explicitly neglected,
the effects of a finite range inherent to the scattering kernel have not been
analyzed for their impact on final-state observables.

It is the goal of this study to ascertain the importance of this second length
scale in high-energy collisions. By varying the interaction range in the
scattering kernel, while leaving the mean free path unchanged, we study the
manisfestations of a non-zero interaction range. We demonstrate that a finite
interaction range contributes to viscous terms in a manner similar to the
finite mean free path, but with different dependencies with respect to density
and temperature. We find that the finite range of the interaction affects the
evolution of heavy-ion reactions, and alters final-state observables,
especially the elliptic flow.

In the next section we present a formal review of the Boltzmann equation and
show how viscosities arise from the interaction range. In section
\ref{sec:spectraflow} we show how spectra and flow observables are sensitive to
the finite range. We discuss algorithmic sensitivities which cannot be ignored
when the interaction range is non-zero, and compare results compiled from four
similar numerical implementations of the Boltzmann equation to illustrate this
sensitivity. We show how viscous heating can be understood from analyzing the
collision kernel in Sec. \ref{sec:collkernel}. In particular we present a
comparison of heating derived from analysis of the collision kernel with the
heating observed in a simple simulation. A discussion of the importance of
non-local interactions and the associated causality problems is provided in the
summary.

\section{Connecting Viscosities to Finite-Range Interactions}
\label{sec:formal}

\subsection{The Role of the Collision Kernel in Boltzmann Descriptions}

The Boltzmann equation can be expressed,
\begin{eqnarray}
\frac{\partial}{\partial t}f({\bf p},{\bf r},t)+
{\bf v}_p\cdot\nabla f({\bf p},{\bf r},t)+{\bf F}({\bf r},t)\cdot \nabla_p
f({\bf p},{\bf r.},t)
&=&\int d^3q d^3q^\prime d^3p^\prime d^3r^\prime dt^\prime\\
&&\hspace*{-200pt}
\cdot\left\{
f({\bf q},{\bf r},t)f({\bf q}^\prime,{\bf r}^\prime,t^\prime)
{\cal}K({\bf r}-{\bf r}^\prime,t-t^\prime;{\bf q},{\bf q^\prime};
{\bf p},{\bf p}^\prime)
-f({\bf p},{\bf r},t)f({\bf p}^\prime,{\bf r}^\prime,t^\prime)
{\cal}K({\bf r}-{\bf r}^\prime,t-t^\prime;{\bf p},{\bf p^\prime};
{\bf q},{\bf q}^\prime)
\right\},
\end{eqnarray}
where $f$ is the phase space density and ${\bf F}$ is the force $d{\bf p}/dt$
felt by a particle at position ${\bf r}$. The collision kernel ${\cal K}({\bf
r}-{\bf r}^\prime,t-t^\prime;{\bf q},{\bf q}^\prime;{\bf p},{\bf p}^\prime)$
describes the differential probability for scattering a pair of particles
separated in space-time by $x-x^\prime$ with initial momenta ${\bf q}$ and
${\bf q}^\prime$ into final states with momenta ${\bf p}$ and ${\bf
p}^\prime$. The range of the collision kernel in coordinate space is the
subject of this paper.

Integrating over the collision kernel should yield the cross section,
\begin{equation}
\label{eq:kernel}
\int d^3r^\prime dt^\prime
K({\bf r}-{\bf r}^\prime,t-t^\prime;{\bf q},{\bf q}^\prime;
{\bf p},{\bf p}^\prime)
=\frac{1}{(2\pi)^3}\frac{d\sigma}{d^3p_{\rm rel}}
v_{\rm rel}\delta^3({\bf p}+{\bf p}^\prime-{\bf q}-{\bf q}^\prime),
\end{equation}
where ${\bf p}_{\rm rel}$ is the relative momentum of the outgoing pair $({\bf
p}-{\bf p}^\prime)/2$.  By inspection of Eq. (\ref{eq:kernel}), one can see
that the coordinate-space dependence of ${\cal K}$ appears rather arbitrary as
long as it integrates to the free-space cross section. Indeed, results at low
density, where particles interact only pairwise, are unaffected by the form of
${\cal K}$ as long as the range is much less than the mean free path and much
less than the characteristic dimensions of the reaction volume.

Any non-zero extent of the collision kernel leads to problems with
super-luminar transport. However, these problems are easily defeated by
restricting the kernel to being local, i.e.,
\begin{equation}
K({\bf r}-{\bf r}^\prime,t-t^\prime;{\bf q},{\bf q}^\prime;
{\bf p},{\bf p}^\prime)
=\frac{1}{(2\pi)^3}\delta^3({\bf r}-{\bf r}^\prime)\delta(t-t^\prime)\\
\cdot\frac{d\sigma}{d^3{\bf p}_{\rm rel}}
v_{\rm rel}\delta^3({\bf p}+{\bf p}^\prime-{\bf q}-{\bf q}^\prime).
\end{equation}
The Boltzmann equation can now be written in a manifestly covariant form.
\begin{eqnarray}
\label{eq:covariantboltzmann}
\left(u_{\bf p}^\mu\partial_\mu +F^\mu\frac{\partial}{\partial p^\mu}\right)
f({\bf p},{\bf r},t)
=\frac{1}{(2\pi)^3}\int \frac{d^3q^\prime}{E_q\prime} \frac{d^3q}{E_q}
\cdot&&\left\{
\sqrt{(q\cdot q^\prime)^2-m^4}\frac{d\sigma}{d^3\tilde{\bf p}_{\rm rel}}
f({\bf q},{\bf r},t)f({\bf q}^\prime,{\bf r},t)
\right.\\
\nonumber
&&-\left.\sqrt{(p\cdot p^\prime)^2-m^4}
\frac{d\sigma}{d^3\tilde{\bf q}_{\rm rel}}
f({\bf p},{\bf r},t)f({\bf p}^\prime,{\bf r},t)
\right\}
\end{eqnarray}
Here, $u_{\bf p}$ is the four-velocity of a particle with momentum ${\bf p}$,
$F^\mu$ is the force $dp^\mu/d\tau$, and $\tilde{\bf p}_{\rm rel}$ is the
relative momentum of the outgoing particles in the center of mass.

\subsection{Effective Viscosities from Finite-Range Interactions}
\label{subsec:hydro}

In this section we describe how interaction over a finite range contributes to
the shear viscosity, $\eta$, the bulk viscosity, $\zeta$, and the thermal
conductivity, $\chi$. We relate the range of the interaction to all three
coefficients.  In order to make this connection, we consider two particles
which scatter from one another, separated by a distance ${\bf r}={\bf r}_2-{\bf
r}_1$. Combining this finite separation with the velocity gradient, one sees
that the first particle interacts with particles which have a higher average
energy. By evaluating the rate at which energy is transferred to the first
particle from colliding with more energetic particles, we find an expression
for the rate at which heat is deposited to the region defined by ${\bf
r}_1$. By comparing to analogous expressions from hydrodynamics, we derive
expressions for all three viscous terms in terms of the interaction range,
$r=|{\bf r}_1-{\bf r}_2|$, the density $n$, and the collision rate $\Gamma$.

Choosing a reference frame such that the velocity of bulk matter at the
location of the first particle is zero, the collective velocity at ${\bf r}_2$
is
\begin{equation}
v_{i}=A_{ij}r_j, ~~A_{ij}=\frac{\partial v_i}{\partial r_j}.
\end{equation}

For an elastic collision where two particles of identical mass simultaneously
change their momenta, the radial components of the momenta must be interchanged
by the collision if energy, linear momentum and angular momentum are to be
conserved. Physically, this corresponds to the scattering from the interior or
exterior surface of a hard sphere. The average energy change of the first
particle is then
\begin{equation}
\langle \delta E_1\rangle
=\langle E_{2,r}\rangle-\langle E_{1,r}\rangle =\frac{m}{2}
\langle({\bf v}\cdot \hat{r})^2\rangle.
\end{equation}
The mass term $m$ is not to be taken literally as the mass of the particles,
since the averaging may include factors of the velocity to account for the flux
or it may have a complicated form to accommodate a desired differential cross
section.  For relativistic motion, the mass might incorporate the lateral
motion of the particles. Writing $\langle \delta E_1\rangle$ in terms of $A$,
\begin{equation}
\langle \delta E_1\rangle =\frac{m}{2r^2}\langle (r_iA_{ij}r_j)^2\rangle.
\end{equation}
One can perform the average over the directions of ${\bf r}$ using the
identity, 
\begin{equation}
A_{ij}A_{kl}\langle r_ir_jr_kr_l\rangle
=A_{ij}A_{kl}\frac{r^4}{15}\left(
\delta_{ij}\delta_{kl}+\delta_{ik}\delta_{jl}+\delta_{il}\delta_{jk}
\right).
\end{equation}
One can then express $\langle \delta E_1\rangle$ in terms of $A$ and $r$.
\begin{eqnarray}
\langle \delta E_1\rangle&=&
\frac{mr^2}{30}\left(
({\rm Tr}A)^2+\frac{1}{2}\sum_{ij}(A_{ij}+A_{ji})^2
\right)\\
\nonumber
&=&\frac{mr^2}{30}\left[
(\nabla\cdot v)^2+\frac{1}{2}\sum_{ij}\left(\frac{\partial v_i}{\partial r_j}
+\frac{\partial v_j}{\partial r_i}\right)^2
\right].
\end{eqnarray}
The rate at which the entropy increases due to these interactions is given by
the density multiplied by the rate at which collisions deposit energy
non-locally, 
\begin{eqnarray}
\label{eq:delsofalpha}
\partial\cdot S&=&\frac{n\Gamma}{T}\langle \delta E_1\rangle\\
\nonumber
&=&\frac{n\Gamma mr^2}{30T}\left[
(\nabla\cdot v)^2+\frac{1}{2}\sum_{ij}\left(\frac{\partial v_i}{\partial r_j}
+\frac{\partial v_j}{\partial r_i}\right)^2
\right].
\end{eqnarray}
Here, $\Gamma$ is the collision rate experienced by a single particle and $n$
is the density.

It is notable that only the symmetric part of $A$ contributes to $\langle\delta
E_1\rangle$. This owes itself to conservation of angular momentum which forbids
rotational motion from being transferred between particles. In fact, if one had
derived an expression for $\langle\delta E_1\rangle$ using $v^2$ instead of
$v_r^2$, the resulting expression would have included the odd parts of $A$
which would correspond to rotational motion, $\nabla\times {\bf v}$. These
terms would have no hydrodynamical analog as they would have reflected a
violation of angular momentum conservation.

We now provide analogous expressions for Eq. (\ref{eq:delsofalpha}) in the
language of hydrodynamics. The expression for entropy production
\cite{weinberghydro} in terms of velocity gradients is
\begin{equation}
\label{eq:weinberg}
\partial\cdot S=
\frac{\eta}{2T}\sum_{ij}\left(\frac{\partial v_i}{\partial x_j}
+\frac{\partial v_j}{\partial x_i}-\frac{2}{3}\delta_{ij}\nabla\cdot{\bf v}
\right)^2
+\frac{\zeta}{T}(\nabla\cdot {\bf v})^2+\frac{\chi}{T^2}(\nabla T)^2.
\end{equation}
One can extract the coefficients by comparing Eq. (\ref{eq:delsofalpha}) to the
first two terms in Eq. (\ref{eq:weinberg}).
\begin{eqnarray}
\label{eq:etazetaresult}
\eta&=&\frac{mn\Gamma r^2}{30}\\
\zeta&=&\frac{mn\Gamma r^2}{18}.
\end{eqnarray}

Applying similar reasoning, one can also derive an expression for the thermal
conductivity. First, we relate the temperature gradient to the energy
flow. Again, we consider particles separated by ${\bf r}$. If collisions occur
between two particles at locations with different temperatures, the average
energy exchanged is
\begin{equation}
\delta E_1=\frac{1}{2}C_r{\bf r}\cdot\nabla T.
\end{equation}
Here $C_r$ represents the change in radial kinetic energy per particle per
change in temperature,
\begin{equation}
C_r=\frac{\partial}{\partial T}E_r.
\end{equation}
In the non-relativistic limit, $C_r=1/2$.

Since the exchange corresponds to moving an energy a finite distance over an
effective time given by the collision rate, one can define the average momentum
density in terms of the energy flow.
\begin{eqnarray}
M_i&=&-\frac{n\Gamma C_r}{4}\langle r_ir_j\rangle\frac{\partial T}{\partial
x_j}.\\
{\bf M}&=&-\frac{n\Gamma C_r r^2}{12}\nabla T
\end{eqnarray}
An extra factor of $1/2$ was added to correct for double counting the
collisions.

One can relate the energy flow to the entropy production,
\begin{eqnarray}
\frac{dS}{dt}&=&\int d^3x\frac{1}{T}\frac{\partial\epsilon}{\partial t}\\
&=&-\int d^3x \frac{1}{T}\nabla\cdot{\bf M}\\
\label{eq:heatingrate}
&=&\int d^3x \frac{n\Gamma C_r r^2}{12T^2}(\nabla T)^2,
\end{eqnarray}
where the continuity equation has been applied.  One can now compare to the
last term in Eq. (\ref{eq:weinberg}) to obtain the thermal conductivity $\chi$.
\begin{equation}
\chi=\frac{n\Gamma C_r r^2}{12}.
\end{equation}

The forms for the three coefficients, $\eta$, $\zeta$ and $\chi$, fundamentally
differ from the forms that result from considering a non-zero mean free path
$\ell$. The viscous coefficients that result from finite $\ell$ are independent
of density and scale inversely with the cross section \cite{weinberghydro}. The
coefficients arising from a non-zero interaction range $r$ scale as the square
of the density and are proportional to the cross section. Thus, non-local
interactions provide viscosities that are important in wholly different
conditions than those where the finite mean free path plays an important
role.  For a rapidly expanding system, finite-range interactions play an
important role when the interaction range multiplied by the velocity gradient
provides velocities of similar magnitudes to local thermal velocities.  Such
conditions exist in the first one fm/c of highly relativistic hadronic
collisions.

The expressions here derive from a very specific picture, non-relativistic
particles moving on straight-line trajectories punctuated by sharp collisions
when the separation equals $r$. However, all interaction at a finite distance
should result in viscous behavior. Relating the distance $r$ to the cross
section might involve a detailed microscopic evaluation of the collision
kernel. This is especially true for relativistic motion. Despite the
complications, one indeed expects the distance $r^2$ to be of similar magnitude
to collision cross sections. It would be interesting to discern whether the
ratio between the shear and bulk viscosities varies for different scattering
models.

\section{Spectra and Flow}
\label{sec:spectraflow}

\subsection{Sampling Factors and Scattering Algorithms}

Numerical realization of the Boltzmann equation is usually accomplished by
performing a classical simulation of the event and applying a large
over-sampling factor $\lambda$ \cite{cywong,bertschdasgupta}. For instance, if
one is sampling a system with 100 particles the Boltzmann simulation might
simulate the trajectories of 3200 particles which would correspond to
$\lambda=32$. The scattering kernel must be diminished by a factor of
$1/\lambda$ to ensure that the scattering rate and mean free path is
independent of $\lambda$. As the Boltzmann equation is built on the hypothesis
that the evolution is completely described by the one-body distribution, the
Boltzmann limit is realized for large $\lambda$ as the correlations between
particles becomes negligible.

The most common way in which the scattering rates are scaled by $1/\lambda$ is
through reducing the cross section. Usually, particles that pass within a
distance $\sqrt{\sigma/\pi}$ of one another are scattered. Scaling the cross
section by a factor of $1/\lambda$ leaves the collision rate and mean free path
unchanged, and in the limit of large $\lambda$ leads to a purely local
scattering kernel which solves numerous problems associated with non-causal
transmission of information and invariance to Loretz boosts.

Although one can solve the covariance problems by choosing a large $\lambda$,
one must then understand whether ignoring the finite interaction range has
affected the results. More realistically, particles do interact at a finite
range through mutually coupling to classical fields or through the quantum
exchange of particles. A variety of ideas have been discussed in the literature
for including such interactions in ways which preserve covariance and are
consistent with the uncertainty principle \cite{browndanielewicz}. However, due
to an assortment of technical challenges, all such approaches remain in the
development stage. Since most comparisons with experiment continue to be
performed with Boltzmann approaches and since understanding the effect of
finite interaction will be more clear with a simple model, we begin our study
of finite-range effects by analyzing classical Boltzmann treatments.

For the investigations described in this subsection, we will always associate
large sampling factors with a reduction in the effective range of the
interaction. However, that is not necessarily the case. Instead of reducing the
spatial extent of the cross section, one could reduce the probability of
scattering by a factor of $1/\lambda$, i.e. not all particles passing within a
distance $\sqrt{\sigma/\pi}$ of one another would scatter. For high sampling
rates collision-finding meshes with granularities determined by the interaction
range are necessary for efficient numerical algorithms. By reducing the cross
section, instead of reducing the scattering probability, one is able to reduce
the mesh size which results in significantly faster execution.

As mentioned above, particles typically scatter in Boltzmann algorithms when
they pass within $\sqrt{\sigma/\pi}$ of one another, at a time chosen to
correspond to the point of closest approach. Aside from choosing the range of
the collision kernel, many other aspects of such algorithms are arbitrary.  For
instance, one might expect that particles should scatter with an
impact-parameter-dependent probability that has a more complicated form than a
simple step function. Even in the non-relativistic limit, scattering at the
point of closest approach violates conservation of angular momentum. In some
algorithms the reaction plane is preserved in each scattering which has been
shown to affect resulting flows in the cascade limit ($\lambda=1$)
\cite{dkahana}. Finally, scatterings need not occur instantly. By delaying the
particles a certain time before they emerge with their asymptotic momenta, one
can effectively alter the pressure. One must coordinate the time delays with
in-medium modifications and knowledge of the phase shifts in order to be
ergodically consistent with thermodynamic properties at the level of the second
virial coefficient \cite{danpratt}.

In the cascade limit, all scattering algorithms carry a degree of arbitrariness
due to issues involving Lorentz invariance
\cite{kortemeyer,zhang_framedependence}. One problem deals with the ordering of
collisions in time. Each collision can be assigned a point in space time. Since
many points would have space-like separations, time ordering is frame
dependent. Finally, since each collision involves changing trajectories over a
finite separation (when $\lambda \ne\infty$) the final outcome can be affected
by the time-ordering prescription.

A second aspect of arbitrariness deriving from relativistic considerations
derives with what is meant by ``point of closest approach''.  Since altering
the two trajectories simultaneously is inherently frame dependent one must
choose a prescription for the time at which each particle is to change course
due to the collision. To illustrate this we consider two particles with momenta
$p_1$ and $p_2$ at space-time coordinates $x_1$ and $x_2$. One can solve for
the points on their trajectories, $x_1^\prime$ and $x_2^\prime$, where they
would reach the point of closest approach as measured by an observer in the
two-particle center-of-mass, by solving for the point where the relative
separation becomes perpendicular to the relative momentum,
\begin{eqnarray}
\label{eq:simultaneousa}
x_1^{\prime,\mu}&=&x_1^\mu+p_1^\mu\frac{P\cdot (X-x_1)}{p_1\cdot P}\\
\label{eq:simultaneousb}
x_2^{\prime,\mu}&=&x_1^\mu+p_2^\mu\frac{P\cdot (X-x_2)}{p_2\cdot P}\\
\label{eq:simultaneousc}
0&=&(x_1^\prime-x_2^\prime)\cdot (q-P\frac{P\cdot q}{P^2}).
\end{eqnarray}
Here $P$ and $q$ are the total and relative momenta respectively, and $P\cdot
X/\sqrt{s}$ is the time at which the collision occurs as measured by an
observer in the center-of-mass frame.  One can solve Eq.s
(\ref{eq:simultaneousa}-\ref{eq:simultaneousc}) for $P\cdot X$, then inserting
into Eq.s (\ref{eq:simultaneousa}) and (\ref{eq:simultaneousb}) one can solve
for $x_1^\prime$ and $x_2^\prime$.

An algorithmic choice that would retain covariance at the two-body level would
be to alter the two momenta at $x_1^\prime$ and $x_2^\prime$ even though the
times would not be simultaneous in other frames. However, in a dense medium
particles may be involved in several collisions which may result in a variety
of events being scheduled between the times $t_1^\prime$ and $t_2^\prime$. In
fact, a particle might be assigned a collision time that precedes its
creation. Additionally, Boltzmann algorithms often incorporate
collision-finding or mean-field meshes. Prescriptions where the two scattered
particles have their trajectories altered at different times become problematic
when coordinating the evolution with the mesh. The most common algorithmic
choice involves choosing a fixed frame to define collision times, then altering
the trajectories at the point of closest approach as measured in that frame.
This prescription becomes questionable when collective velocities become
relativistic. Such is the case at RHIC where particle emission is spread over
10 units of rapidity. This problem can be somewhat alleviated by ordering the
collisions according to $\tau=(t^2-z^2)^{1/2}$. This choice seems reasonable
for ultra-relativistic collisions where conditions are determined largely by
$\tau$, but seems unreasonable for lower energy collisions where longitudinal
boost invariance is not realized.

\subsection{Comparing Results from Four Algorithms}
\label{subsec:pionwind}

To illustrate the sensitivity of results to algorithmic choices, we compare
results from four numerical approximations to the Boltzmann equation which are
realized in four separate models, each arising from a different author.  All
the algorithms discussed here scatter particles instantaneously at the point of
closest approach, and all the models assume a simple $s-$wave form to the cross
section. Each algorithm was executed from the same initial conditions, and each
was executed both in the cascade limit ($\lambda=1$) and with a high sampling
factor, $\lambda=16$. Due to the arbitrariness inherent to algorithms with
finite-range interactions it should not be surprising for results to vary
between codes with $\lambda=1$, but it is expected that results from the four
approaches should be indistinguishable in the large $\lambda$ limit.

The four models differ in their definition of ``simultaneous'', in their
time-ordering prescriptions and in whether they preserve the scattering plane
in two-particle scattering. The four algorithms compared here represent the
following choices:

\begin{enumerate}
\item ZPC: A parton cascade code authored by
B. Zhang\cite{zpc_longwriteup}. Particles scatter at the point of
closest approach, simultaneous in a fixed reference frame. This time is
determined by first finding the space-time points $x_1^\prime$ and
$x_2^\prime$, where the two particles would scatter should they alter their
trajectories simultaneously in the two-particle center-of-mass frame. In a
designated laboratory frame, a time is determined by averaging $t_1^\prime$ and
$t_2^\prime$. The trajectories of both particles are simultaneously altered at
this average time.  The reaction plane is preserved in each two-body
scattering, and rescattering between the same partners is allowed after one
particle has rescattered.
\item GROMIT-$t$: A generic scattering engine developed for the RHIC
Transport Theory Collaboration (RTTC)\cite{rttc}. This should be identical to
ZPC. 
\item MPC(0.4.0): A Boltzmann description authored by D. Moln\'ar
\cite{mpc}. The principal difference with the codes above is that the reaction
plane is not preserved in the two-body scattering. Another difference is that
rescattering between collision partners is not allowed until both particles
have rescattered. It should be noted that MPC has switches which make it
possible to reproduce the choices described in ZPC and GROMIT-$t$.
\item GROMIT-$\tau$: Another version of cascade/Boltzmann engine developed by
the RTTC collaboration with collisions ordered by $\tau$ to be appropriate for
ultra-relativistic collisions \cite{rttc}. The space-time points at which the
two particles would scatter had the particles scattered simultaneously in the
two-particle rest frame, $x_1^\prime$ and $x_2^\prime$, are used to generate
two proper times, $\tau_1^\prime$ and $\tau_2^\prime$. The average
$\tau^\prime=(\tau_1^\prime+\tau_2^\prime)/2$ is then used for ordering. Both
trajectories are altered when their proper time equals $\tau^\prime$. The
reaction plane is preserved in the scattering and scattering between the same
pair of particles is allowed after one of the particles has rescattered.
\end{enumerate}

Calculations were performed for all four models with identical initial
conditions based on a thermal Bjorken geometry \cite{bjorken}. In Bjorken
coordinates, the time is replaced by a proper time,
\begin{equation}
\tau\equiv\sqrt{t^2-z^2},
\end{equation}
and the position along the beam axis, $z$, is replaced by the effective
coordinate, 
\begin{equation}
\eta\equiv \frac{1}{2}\log\left(\frac{t+x}{t-z}\right).
\end{equation}
The coordinate $\eta$ also gives the rapidity of an observer moving at constant
velocity from the origin at $t=0$ to the space-time point $(t,z)$. 

Thermal emission from a Bjorken geometry would be realized by choosing thermal
sources with positions equally distributed in $\eta$, with source rapidities
$y_s=\eta$, and with emission taking place at a fixed proper time, $\tau$. Such
emission would appear invariant to boosts along the beam axis. In the
calculations described here, the sources are confined to a finite range in
$\eta$,
\begin{equation}
-2<\eta<2.
\end{equation}
The transverse spatial coordinates were randomly placed in a cylinder of radius
5 fm. The simulation involved 2400 pions and 240 protons which were assigned
momenta according to a thermal distribution with a temperature $T=180$ MeV for
the protons and $T=165$ MeV for the pions, to roughly acheive consistency with
spectra resulting from heavy ion collisions. The initial particle densities
were 7.64 pions and 0.764 protons per fm$^3$. To be as simple as possible,
constant cross sections were imposed independent of the species involved.

\begin{figure}
\centerline{\epsfxsize=0.45\textwidth
\epsfbox{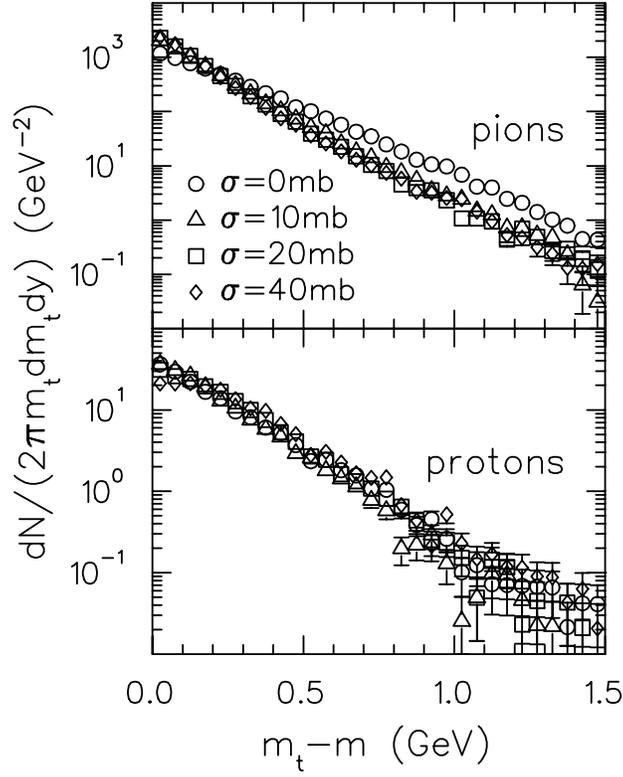}}
\caption{Spectra from GROMIT-$\tau$ are displayed for four cross sections.
\label{fig:pionwind-sigcompare}}
\end{figure}

The resulting spectra at mid-rapidity as calculated with GROMIT-$\tau$ are
shown in Fig. \ref{fig:pionwind-sigcompare} for four different cross sections,
0, 10, 20 and 40 mb. These calculations were performed with a sampling
factor $\lambda=32$. For larger cross sections, the pion spectra are cooler
while the proton spectra are somewhat hotter. The difference would be more
pronounced had the pion and proton distributions been initialized with
identical temperatures.

\begin{figure}
\centerline{\epsfxsize=0.45\textwidth
\epsfbox{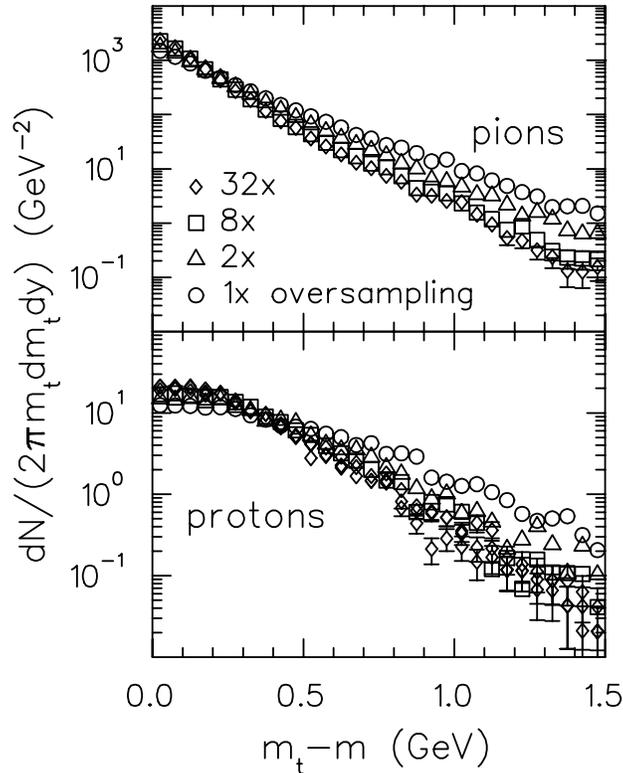}}
\caption{Spectra for pions and protons resulting from GROMIT-$\tau$ with 40 mb
cross sections for four sampling factors, $\lambda=1,2,8,32$.
\label{fig:pionwind-lamcompare}}
\end{figure}

Figure \ref{fig:pionwind-lamcompare} displays spectra from GROMIT-$\tau$ with
$\sigma$ =40 mb for four different sampling factors, $\lambda=$ 1, 2, 8 and
32. The sensitivity to $\lambda$ is remarkable. For small $\lambda$ it appears
that both the pion and proton spectra become hotter. Results appear to converge
for $\lambda\rightarrow\infty$ as the $\lambda=8$ and $\lambda=32$ results are
barely distinguishable. In all the models investigated here, the final-state
transverse energies were higher for small sampling factors. We attribute this
reduction in cooling to viscous effects arising from finite-range interactions.

\begin{figure}
\centerline{\epsfxsize=0.45\textwidth
\epsfbox{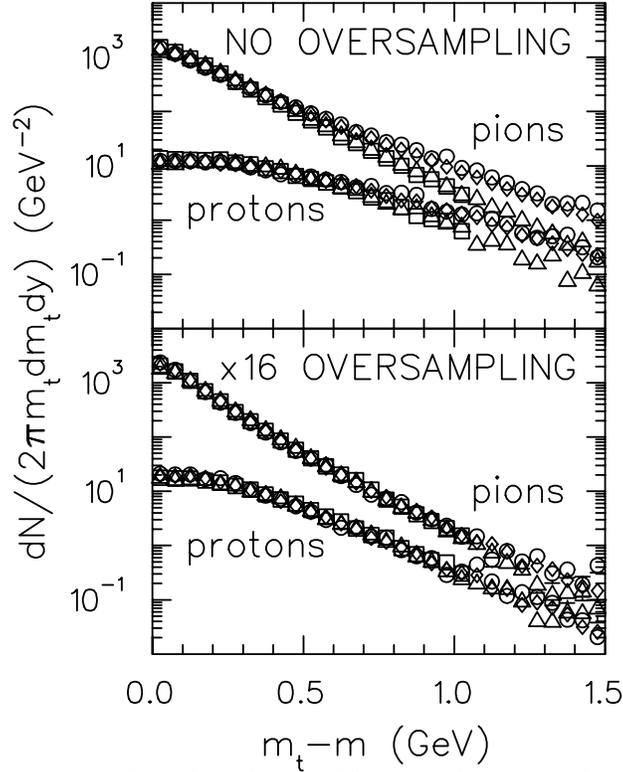}}
\caption{Spectra for pions and protons resulting from four models run with
$\sigma=40$ mb: (ZPC -- squares, MPC -- diamonds, GROMIT-t -- triangles,
GROMIT-$\tau$ -- circles). In the upper panel calculations were performed with
a sampling factor of 1, while in the lower panel each model used a
sampling factor $\lambda=16$. Due to different scattering algorithms, the
models generated different results for $\lambda=1$, while generating identical
results for larger $\lambda$.
\label{fig:pionwind-modcompare}}
\end{figure}

In the upper panel of Fig. \ref{fig:pionwind-modcompare} the resulting spectra
from the four algorithms are displayed in the cascade limit. The slopes vary by
approximately 20 MeV. The same results are displayed in the lower panel of
Fig. \ref{fig:pionwind-modcompare} but with a high sampling factor,
$\lambda=16$.  The various algorithms are then in agreement. Since the hadrons
are initially separated by approximately 1/3 fm, while the interaction range is
1.1 fm, it is not surprising that the $\lambda=1$ result is sensitive to the
scattering algorithm.

For each algorithm the spectra fall significantly more steeply when produced
with a higher sampling factor $\lambda$. This is especially true for the
proton spectra, which suggests that radial flow might be stronger in
the cascade limit. Since the average transverse energy for pions is also larger
in the cascade limit, one can infer that longitudinal cooling is significantly
suppressed through particles interacting at a finite range.

\subsection{Elliptic Flow}
\label{subsec:ellipticflow}

As a second exploration of the sensitivities to the sampling factor, we
consider elliptic flow as observed from the GROMIT-$\tau$ algorithm. In this
case cyclic boundary conditions in $\eta$ were employed to simulate a truly
boost-invariant system. The densities of protons and pions were chosen
identically as above, $n_\pi=7.64$ fm$^{-3}$, $n_p=0.764$ fm$^{-3}$ at
$\tau=1.0$ fm/c. The temperatures for the pions and protons were again chosen
to be 165 MeV and 180 MeV respectively. Aside from the cyclic boundary
conditions in $\eta$ the only difference with the initial conditions used in
the previous subsection is that the particles were confined to an ellipse
rather than a circle in the transverse direction. The two radii of the ellipse
were 5 fm and 2.5 fm.

The elliptical shape resulted in elliptic flow which is parameterized by the
observable $v_2$.
\begin{equation}
v_2=\langle \cos(\phi-\phi_0)\rangle,
\end{equation}
where $\phi_0$ specifies the direction of the reaction plane which contains the
short axis of the ellipse. Elliptic flow has been proposed as a means for
determining the equation of state of the matter at early times
\cite{elliptic_ollitrault,elliptic_danielewicz,elliptic_ags,elliptic_na49,elliptic_sorge,elliptic_zpc,elliptic_star}.

\begin{figure}
\centerline{\epsfxsize=0.45\textwidth
\epsfbox{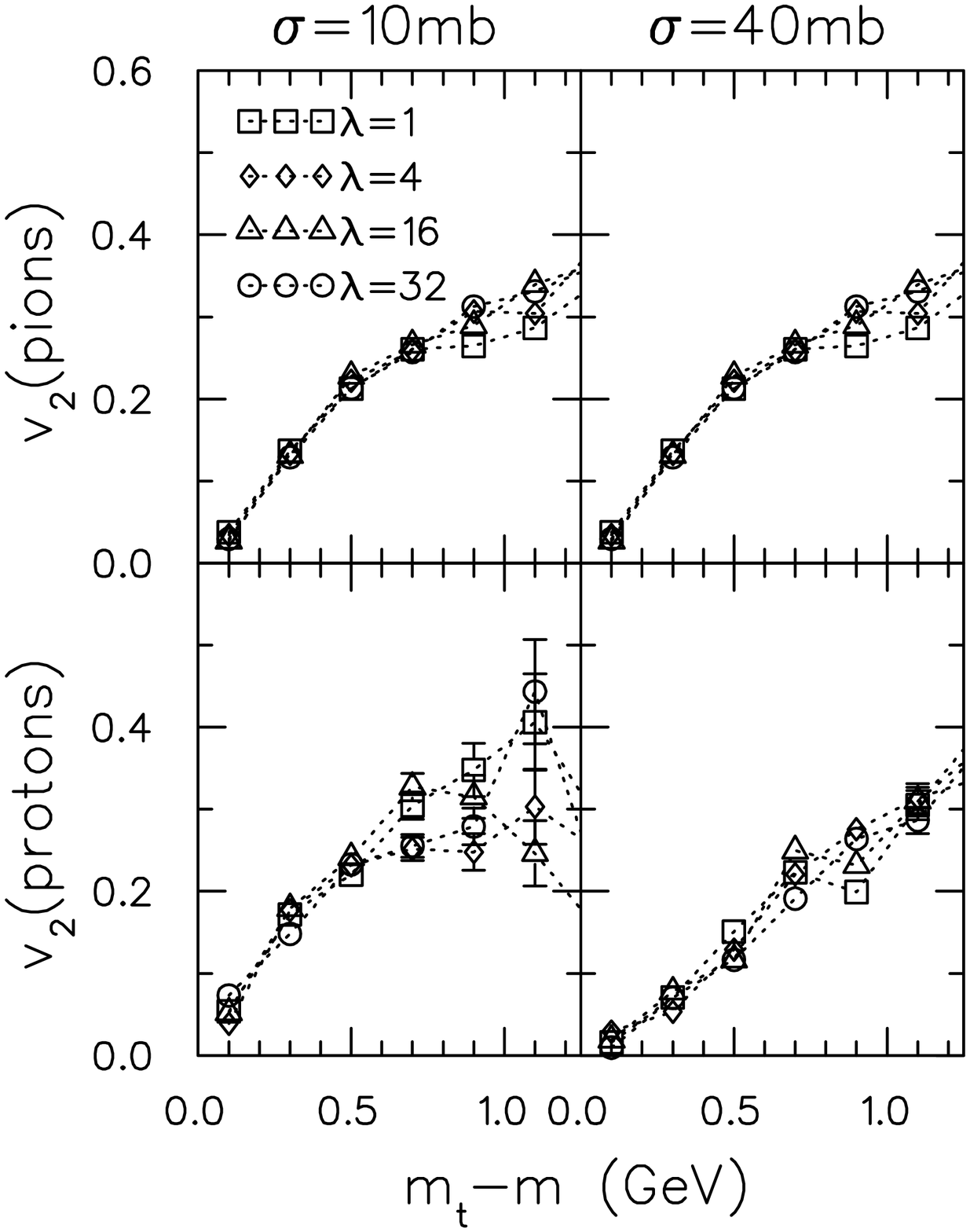}}
\caption{Elliptic flow is shown as a function of transverse energy for both
pions and protons for several sampling factors, $\lambda$. The cross sections
were not varied as $\lambda$ was varied, but instead only a fraction
$1/\lambda$ of the particles were scattered. The insensitivity to $\lambda$
demonstrates that the Boltzmann limit is effectively realized, even for
$\lambda=1$.
\label{fig:ellipticflow_noscale}}
\end{figure}

\begin{figure}
\centerline{\epsfxsize=0.45\textwidth
\epsfbox{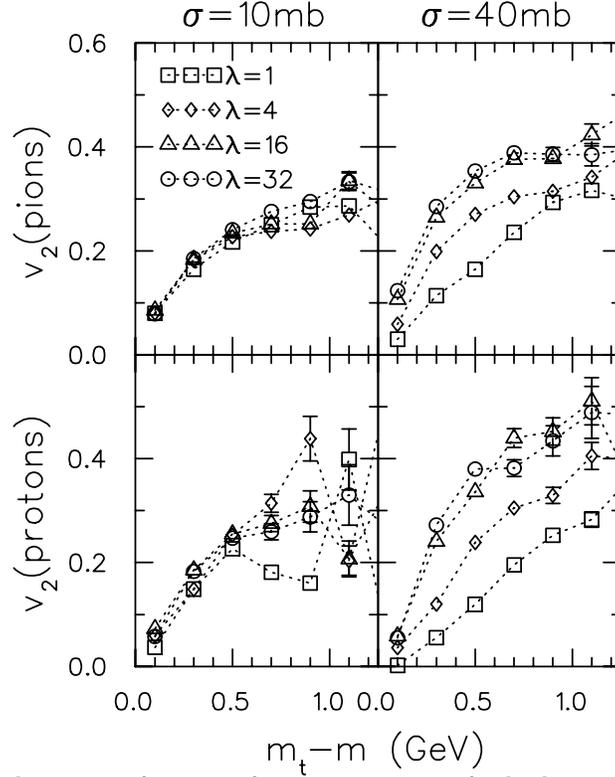}}
\caption{Again, elliptic flow is shown as a function of transverse energy for
both pions and protons for several sampling factors. In this case the cross
sections were scaled by a factor $1/\lambda$ to account for the over-sampling.
The reduced flow resulting for calculations with small $\lambda$ (larger cross
sections) is attributed to viscous effects arising from the finite interaction
range.
\label{fig:ellipticflow}}
\end{figure}

Resulting flows are shown in Fig. \ref{fig:ellipticflow_noscale} and
Fig. \ref{fig:ellipticflow} for two cross sections, 10 mb and 40 mb, and for
four sampling factors.  In Fig. \ref{fig:ellipticflow_noscale} the cross
section was left constant as $\lambda$ was varied. Instead, only a fraction,
$1/\lambda$, of the possible scatterings were executed to account for the
over-sampling. The lack of a sensitivity to $\lambda$ suggests that the
Boltzmann limit is reached even for $\lambda=1$, i.e. $n$-body correlations do
not signficantly affect the flow.  The same quantities were calculated in
Fig. \ref{fig:ellipticflow}, except that the cross sections were scaled as
$1/\lambda$ to account for the over-sampling.  The results are then highly
sensitive to $\lambda$ which suggests that the finite interaction range that
accompanies small $\lambda$ leads to a reduced elliptic flow.

The reduction in longitudinal work and the damping of elliptic flow due to the
incorporation of a finite-range interaction into a Boltzmann treatment suggests
that viscous effects have effectively been introduced into the reaction.

\section{Analyzing the Collision Kernel}
\label{sec:collkernel}

In Sec. \ref{subsec:hydro} a finite range of interaction, characterized by a
length scale, $r$, was shown to generate viscous behavior. This length scale is
determined by the cross section, $r^2\sim\sigma/\pi$, but the constant of
proportionality is not trivially determined and can depend on seemingly
arbitrary aspects of scattering algorithms. In the following subsection, we
illustrate how viscous heating can also be directly related to the collision
kernel by considering a simple example of one-dimensional expansion with a
Bjorken space-time geometry. In the subsequent subsection we compare
predictions based on the form of the scattering kernel with results from
simulations based on the same kernel.

\subsection{Viscous Heating in a Bjorken Expansion}

As a simple example we again consider a one-dimensionally boost-invariant
system of infinite extent in all dimensions. We keep the number of particles
fixed so that the density scales inversely with $\tau$,
\begin{equation}
n(\tau)=\frac{1}{\tau}dN/(dAd\eta).
\end{equation}
where the number of particles per area per rapidity interval, $dN/(dAd\eta)$,
is fixed. To simplify matters we consider massless particles.

Here we wish to calculate the rate at which a particle at $\eta=0$ has
collisions with other particles, $dN_c/d\tau$, and the rate at which it gains
or loses energy from such collisions, $d\langle E_1\rangle/d\tau$. We assume
the particles are in local thermal equilibrium, and analyze the collision
kernel to determine the rates. 

Refering to the colliding partners with the subscript `2', the collision and
heating rates per particle are
\begin{eqnarray}
\label{eq:kernelrate}
 \frac{dN_c}{d\tau} &=& 2\pi \frac{dN}{dA\, d\eta} \int d\eta_2 \, r_2\, dr_2\,
d^3p_2 \frac{dN}{d^3p_2}\\ \nonumber
&&\hspace*{70pt}\cdot\delta(\tau_c-\tau)\Theta\left(\frac {\sigma}{\pi} -
b^2\right), \\
\label{eq:kernelerate}
\frac{d\langle E_1\rangle}{d\tau} &=&2\pi \frac{dN}{dA\, d\eta}
\int d\eta_2\, r_2\, dr_2 \, d^3p_2 \frac{dN}{d^3p_2}\\ \nonumber
&&\hspace*{60pt}\cdot\delta(\tau_c-\tau)\Theta\left(\frac {\sigma}{\pi} -
b^2\right) \delta E_1.
\end{eqnarray}
Here, $\delta E_1$ is the average change in energy of the particle `1' due to
the collision and $r_2$ and $\eta_2$ describe the position of the second
particle. The impact parameter for the two-particle collision is $b$ and
$\tau_c$ is the collision time.

The strategy employed here is to calculate the collision time, $\tau_c$. The
first particle is at a position $x_1=(\tau,0,0,0)$ with four momentum
$p_1$. The second particle is at a position $x_2=(\tau\cosh\eta_2,
r_2,0,\tau\sinh \eta_2)$, with a four-momentum $p_2$. Once $\tau_c$ is
understood in terms of $r_2$, $\eta_2$, $p_1$ and $p_2$, one can replace the
delta function in the expression above,
\begin{equation}
\delta(\tau_c-\tau)\rightarrow \delta(r_c-r_2)
\frac{\partial r_c}{\partial \tau_c},
\end{equation}
where $r_c$ is the position required to make the collision occur at $\tau$. By
substituting the delta function with $r_c$ for the delta function with
$\tau_c$, the integrals in Eq. (\ref{eq:kernelrate}) and
Eq. (\ref{eq:kernelerate}) can be simplified and solved numerically.

The first step one must perform is to find $\tau_c$ in terms of $r$. The prescription
for finding $\tau_c$ is somewhat arbitrary due to covariance issues. For our
purposes, we will first define a time in the rest frame of the two particles
where the particles run abreast of one another, $t_c$. With $x_c= (t_c,0,0,0)$
in the center-of-mass frame and $P= p_1+ p_2$, the collision time is determined
by the condition,
\begin{eqnarray}
(p_1-p_2)\cdot (x_1^\prime-x_2^\prime)&=&0, \\
x_1^\prime &=& x_1+\frac{P\cdot(x_c-x_1)}{P\cdot p_1} p_1,\\
x_2^\prime &=& x_2+\frac{P\cdot(x_c-x_2)}{P\cdot p_2} p_2.
\end{eqnarray}
One can solve for $P\cdot x_c$,
\begin{equation}
P\cdot x_c= p_1\cdot x_1+p_2\cdot x_2.
\end{equation}
One can now solve for $x_1^\prime$ and $x_2^\prime$.
\begin{eqnarray}
x_1^\prime&=&x_1+p_1\frac{p_2\cdot(x_2-x_1)}{p_1\cdot p_2}\\
x_2^\prime&=&x_2+p_2\frac{p_1\cdot(x_1-x_2)}{p_1\cdot p_2}
\end{eqnarray}
If one calculated the times at which the collision occurred, $t_1^\prime$ and
$t_2^\prime$, they would only be simultaneous in the two-particle
center-of-mass frame. Similarly, if one calculated $\tau_1^\prime =(t_1^{\prime
2}-z_1^{\prime 2})^{1/2}$ and $\tau_2^\prime =(t_2^{\prime 2}-z_2^{\prime
2})^{1/2}$, one would find that the two Bjorken proper times would not be
simultaneous. One must arbitrarily choose a prescription to choose $\tau_c$ in
terms of $x_1^\prime$ and $x_2^\prime$. For our purposes, we choose the
following prescription,
\begin{eqnarray}
\tau_c^2\equiv\frac{\tau_1^{\prime 2}+\tau_2^{\prime 2}} {2}
\end{eqnarray}
Our only motivation in averaging the $\tau^2$s rather than the $\tau$s is that
the algebra simplified. Using this choice,
\begin{eqnarray}
\label{quadrc_eq}
\tau_c^2&=& \tau^2 + \frac{C_0r'^2+2C_1r'+C_2} {2}\\
\nonumber
C_0&=&\gamma_1^2p_{1\perp}^2+\gamma_2^2p_{2\perp}^2\\
\nonumber
C_1&=&\gamma_1\alpha_1+\delta_1\gamma_1p_{1\perp}^2
+\gamma_2\alpha_2+\delta_2\gamma_2p_{2\perp}^2\\
\nonumber
C_2&=&\delta_1^2p_{1\perp}^2+2\delta_1\alpha_1+\delta_2^2p_{2\perp}^2
+2\delta_2\alpha_2
\end{eqnarray}
The coefficients are defined
\begin{eqnarray}
\alpha_1=E_1t-p_{1,z}z~,&&\alpha_2=E_2 t+p_{2,z}z\\
\nonumber
\delta_1=2zp_{2,z}/(p_1\cdot p_2)~,&&\delta_2=-2zp_{1,z}/(p_1\cdot p_2)\\
\nonumber
\gamma_1=2p_{2,x}/(p_1\cdot p_2)~,&&\gamma_2=-2p_{1,x}/(p_1\cdot p_2)
\nonumber
\end{eqnarray}
Here $t$, $r$ and $z$ are the positions describing the first particle in a
frame where the particles are centered about $\eta=0$ and $r=0$.
\begin{equation}
  \eta' = \frac{\eta_2}{2}, ~~t=\tau\cosh\eta',~~z=-\tau\sinh \eta',
  ~~r'=\frac{r_2}{2}.
\end{equation}
In terms of these new variables, one can rewrite the rates above,
\begin{eqnarray}
\frac{dN_c}{d\tau}&=&16\pi\frac{dN}{dA\, d\eta} \int d\eta'\, r'\, dr'\, d^3p_2
\frac{dN}{d^3p_2}\\
&&\nonumber 
\hspace*{80pt}\cdot\delta(\tau-\tau_c)
\Theta\left(\frac {\sigma}{\pi} - b^2\right)\\
\nonumber
&=&16\pi \frac{dN}{dA\, d\eta} \int d\eta'\, r'\, dr'\, d^3p_2
\frac{dN}{d^3p_2} \\
\nonumber 
&&\hspace*{40pt}
\cdot\delta(r^\prime-r_c)\frac{2\tau}{C_0r^\prime+C_1}
\Theta\left(\frac {\sigma}{\pi} - b^2\right) ,
\\
\frac{d\langle E_1\rangle}{d\tau}&=&16\pi \frac{dN}{dA\, d\eta}
\int  d\eta'\, r'\, dr'\, d^3p_2
\frac{dN}{d^3p_2}\\
\nonumber
&&\hspace*{15pt}
\cdot\delta(r^\prime-r_c)\frac{2\tau}{C_0r^\prime+C_1} 
\Theta\left(\frac {\sigma}{\pi} - b^2\right) 
\delta E_1.
\end{eqnarray}
When solving for $r_c$, there are two solutions to Eq. (\ref{quadrc_eq}). If
one uses solutions for both positive and negative $r_c$, the factor $16\pi$
should be reduced to $8\pi$.

One can express these results as averages over $p_1$ and $p_2$
\begin{eqnarray}
\frac{dN_c}{d\tau}&=&32\pi\tau\frac{dN}{dA\, d\eta} \int d\eta' 
\left\langle\frac{r_c}{C_0r_c+C_1}\right\rangle\\
\nonumber
&&\hspace*{100pt}\cdot\Theta\left(\frac {\sigma}{\pi} - b^2\right),\\
\label{eq:collkernel}
\frac{d\langle E_1\rangle}{d\tau}&=&32\pi\tau\frac{dN}{dA\, d\eta} \int d\eta' 
\left\langle\frac{r_c}{C_0r_c+C_1}
\delta E_1 \right\rangle\\
\nonumber
&&\hspace*{100pt}\Theta\left(\frac {\sigma}{\pi} - b^2\right).
\end{eqnarray}
If scattering angles are chosen with equal probability forwards and backwards,
the average change in the two energies is
\begin{equation}
\delta E_1=\frac{E_2-E_1}{2}.
\end{equation}

The non-local aspect of the collision kernel should contribute to heating the
particles in their local frame. From physical arguments, one expects the
non-local contribution to heating to scale with temperature, density, time and
cross section in a simple manner.
\begin{eqnarray}
\label{eq:nonlocaleperp}
\frac{dE_\perp}{d\tau}&=&\frac{3}{4\pi}\frac{d\langle E_1\rangle}{d\tau}\\
\nonumber
&=&\beta\frac{dN}{dAd\eta}\frac{\sigma^2T}{\tau^3},
\end{eqnarray}
where $E_t$ is the mean transverse energy per particle and $\beta$ is a
dimensionless constant.

The simple scaling derives from the hydrodynamically motivated Eq.s
(\ref{eq:weinberg}) and (\ref{eq:etazetaresult}). One expects the collision
rate to scale proportional to the density, which requires the factor
$dN/(dAd\eta)$ and one inverse power of $\tau$, and the cross-section. The
squared-velocity gradient suggests an extra factor of $\tau^{-2}$, and the
range of the interaction requires an extra factor of $\sigma$.  The constant
$\beta$ is determined by the form of the differential cross section,
e.g., $s$-wave scattering would result in a higher $\beta$ than a highly
forward-peaked form. Since one power of $\sigma$ comes from the range of
interaction, $\beta$ should scale inversely with the sampling factor
$\lambda$.  Finally, the effective mass should be proportional to the
temperature.

The heating rate due to non-local interactions is calculated from
Eq. (\ref{eq:collkernel}) by numerically analyzing the collision kernel and is
displayed in Fig. \ref{fig:eperpscaling} after being scaled by the temperature,
time and cross section. Had the simple scaling argument been correct the ratio
would have been a constant $\beta$. However, due to higher order corrections
in $1/\tau$, the ratio varies as a function of $\tau$. The ratio approaches a
constant for large $\tau$, $\tau^2>>\sigma$. The scaling fails when $\tau$
becomes less than the range of the interaction, which for this example is $\sim
1.0$ fm due to the 10 mb cross section. The departure of the ratio from a
straight line illustrates the limitation of simple hydrodynamic arguments to
describe the viscous heating from non-local interactions.

\subsection{Comparing to Numerical Results}

Finally, we present numerical results involving a boost-invariant Boltzmann
description (GROMIT-$\tau$) which was executed with cyclic boundary conditions,
both in $\eta$ and in the transverse coordinates. We compare the viscous
heating rate observed in the numerical calculation with the viscous heating
rate expected from the scaling arguments expressed in
Eq. (\ref{eq:nonlocaleperp}), where the coefficient $\beta$ was determined
from analyzing the collision kernel from the last section. The temperature was
set to 400 MeV at a time $\tau=0.1$ fm/c, and the cross section was chosen to
be 10 mb. A simple $s$-wave form was used for the angular dependence of the
cross section.

\begin{figure}
\centerline{\epsfxsize=0.45\textwidth
\epsfbox{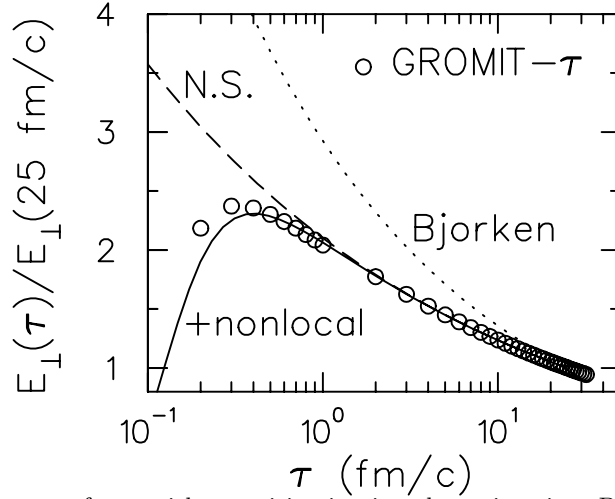}}
\caption{The mean transverse energy for particles participating in a
boost-invariant Bjorken expansion is displayed as a function of the proper time
(circles). Also displayed are the Bjorken hydrodynamic result (dotted line),
the Navier-Stokes correction which accounts for viscous shear arising from a
finite mean free path (dashed line) and the correction due to non-local
interactions as expected from simple scaling arguments (solid line).  Non-local
interactions are important at small times when the velocity gradients and
collision rates are high.
\label{fig:swaveresult}}
\end{figure}

The resulting mean transverse energy is displayed in Fig. \ref{fig:swaveresult}
as a function of $\tau$. The initial heating derives from the non-local nature
of the interactions. Longitudinal cooling ultimately dominates the behavior as
the non-local contribution to the heating falls roughly at $\tau^{-3}$.

The dotted line in Fig. \ref{fig:swaveresult} describes the evolution of the
transverse energy in the limit of ideal (non-viscous) hydrodynamics. In that
limit, the stress-energy tensor has a simple form,
\begin{equation}
T^{\alpha\beta}=\epsilon u^\alpha u^\beta+P(u^\alpha u^\beta-g^{\alpha\beta}).
\end{equation}
In the Bjorken limit, $\partial v/\partial z=1/\tau$ and the evolution of the
energy density becomes
\begin{equation}
\frac{\partial}{\partial t}\epsilon=-\frac{P+\epsilon}{\tau}.
\end{equation}
For the massless case, $P=\epsilon/3$, which gives the result,
\begin{equation}
\epsilon(\tau)=\epsilon(\tau_0)\left(\frac{\tau}{\tau_0}\right)^{-4/3}
\end{equation}

The velocity gradient generates a shear which contributes an additional term
to the stress-energy tensor.
\begin{equation}
T_{\eta}^{\alpha\beta}=\eta\left(-g^{\alpha\gamma}-u^\alpha
u^\gamma\right) \left(\frac{\partial u_\gamma}{\partial x_\beta}
+\frac{\partial u_\beta}{\partial x_\gamma}-\frac{2}{3}g_\gamma^\beta
\partial\cdot u\right).
\end{equation}
In a simulation, one can determine $\eta_{NS}$ by evaluating the stress-energy
tensor.
\begin{equation}
\frac{T^{xx}+T_{yy}}{2}-T^{zz}=2 \frac{\eta_{NS}}{\tau}.
\end{equation}
The components of the stress-energy tensor are extracted by analyzing the
momenta of particles as measured in the local rest frame.
\begin{equation}
T_{ij}=\frac{1}{V}\sum \frac{p_ip_j}{E_p}.
\end{equation}

The Navier-Stokes evolution of the energy density is governed by the equation,
\begin{equation}
\frac{\partial}{\partial \tau}\epsilon=-\frac{P+\epsilon}{\tau}
+\frac{4}{3}\frac{\eta_{NS}}{\tau^2}.
\end{equation}
For massless particles interacting with a constant cross section, dimensional
arguments force the viscosity to rise linearly with $\tau$ since the mean free
path should grow with $\tau$ due to the density falling as $1/\tau$.
\begin{equation}
\eta_{NS}=C_{NS}\epsilon\tau,
\end{equation}
where $C_{NS}$ is a constant, determined only by the form of the cross
section. The energy density then follows the Navier Stokes form~
\cite{zpc-navstokes}. This form should be valid unless the viscous contribution
to the stress energy tensor approaches the same order as the equilibrated
pressure \cite{danielewiczgyulassy}.
\begin{equation}
\epsilon(\tau)=\epsilon(\tau)\left(\frac{\tau}{\tau_0}\right)^{-(4/3)
(1-C_{NS})}.
\end{equation}
The Navier Stokes result is represented by a dashed line in
Fig. \ref{fig:swaveresult}. The value of $C_{NS}$ was determined by
evaluating the asymmetry of the stress-energy tensor in the simulation at large
times. Running simulations with a large sampling ratios generated results
in excellent agreement with the Navier Stokes result.

\begin{figure}
\centerline{\epsfxsize=0.45\textwidth
\epsfbox{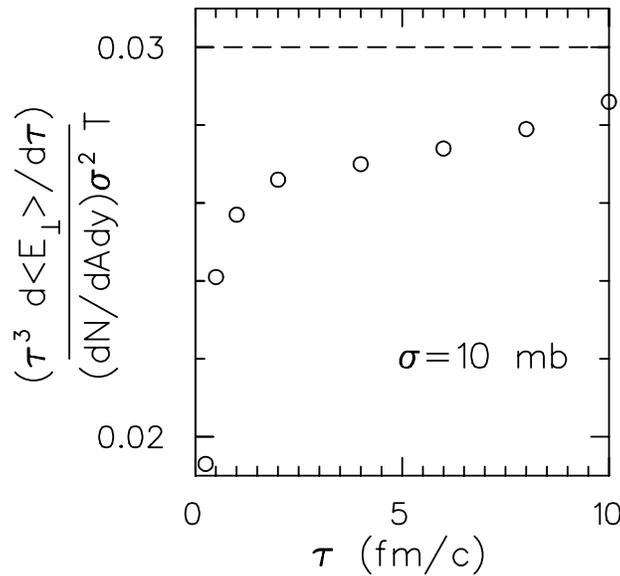}}
\caption{The heating due to non-local interactions as calculated numerically
from the collision kernel (circles) is scaled in such a way that it would be
constant if the simple hydrodynamic scaling arguments were valid. The dashed
line represents the asymptotic value.
\label{fig:eperpscaling}}
\end{figure}

The inclusion of non-local effects is responsible for the discrepancy
between the simulation results in Fig. \ref{fig:swaveresult} and the
Navier-Stokes results. Eq. (\ref{eq:etazetaresult}) suggests that the non-local
correction to the viscosity should scale proportional to $\tau^{-2}$,
\begin{equation}
\eta_{nl}=C_{nl}\epsilon/\tau,
\end{equation}
where $C_{nl}$ is independent of $\tau$ and scales with $\sigma^2$ as explained
in the previous section. One can now determine the evolution of the energy
density by solving the hydrodynamical equations of motion,
\begin{eqnarray}
\frac{\partial}{\partial \tau}\epsilon&=&
-\frac{4}{3}(1-C_{NS})\frac{\epsilon}{\tau}
+\frac{4}{3}C_{nl}\frac{\epsilon}{\tau^3}\\
\end{eqnarray}
This equation has a simple solution.
\begin{equation}
\label{eq:non-localepsilon}
\epsilon(\tau)
=\epsilon(\tau_0)\left(\frac{\tau}{\tau_0}\right)^{4/3(1-C_{NS})}
\exp\left\{\frac{2C_{nl}}{3}\left(\frac{1}{\tau_0^2}-\frac{1}{\tau^2}\right)
\right\}.
\end{equation}
This form is shown with the solid lines in Fig. \ref{fig:swaveresult}. The
constant $C_{nl}$ was determined by the asymptotic limit of
Fig. \ref{fig:eperpscaling}. The effect of non-local interactions is somewhat
overestimated at small times by Eq. (\ref{eq:non-localepsilon}) as would be
expected by considering Fig. \ref{fig:eperpscaling} which shows that that
growth of the viscous heating at small $\tau$ is somewhat slower than the naive
expectation that it scales as $\tau^{-3}$.

\begin{figure}
\centerline{\epsfxsize=0.45\textwidth
\epsfbox{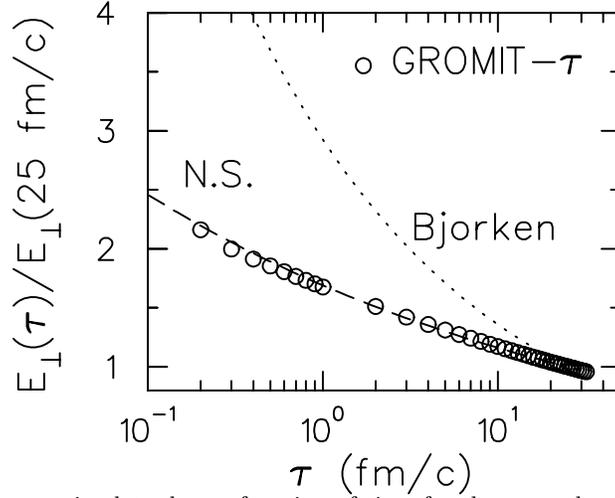}}
\caption{The mean transverse energy is plotted as a function of time for the
case where a forward-peaked cross section as described in
Eq. (\ref{eq:forwardpeaked}) is implemented. The results from GROMIT-$\tau$
(circles) are well described by the Navier-Stokes (dashed line) correction to
the Bjorken solution (dotted line). As compared to $s$-wave scattering, the
Navier-Stokes viscosity is increased while the non-local contribution to the
viscosity becomes negligible.
\label{fig:qcdresult}}
\end{figure}

The effect of non-local interactions is lessened for scatterings that are more
forward peaked. Fig. \ref{fig:qcdresult} illustrates the behavior of $E_t$ as
a function of $\tau$ for a screened Rutherford scattering,
\begin{eqnarray}
\label{eq:forwardpeaked}
\frac{d\sigma}{dt}&=&\frac{9\pi}{2}\frac{\alpha_s^2}{(t-\mu^2)^2},\\
\sigma&=&\frac{9\pi\alpha_s^2}{2\mu^2},\\
t&=&(p_1-p_3)^2.
\end{eqnarray}
Here, $\alpha_s=\sqrt{2/9}$, and the screening mass $\mu$ is chosen to provide
a cross section of 10 mb.  As compared to the $s$-wave
scattering result in Fig. \ref{fig:swaveresult}, the effect of non-local
interactions is diminished while the Navier-Stokes viscosity is increased. This
is expected since the mean free path is effectively increased while the energy
transfer inherent to collisions is decreased.

\section{Discussion and Summary}
\label{sec:conclusions}

The effects of short-range, but non-local, interactions provide an unwelcome
complication to the analysis and interpretation of experimental results from
relativistic heavy ion collisions. Flow observables are most significantly
affected, especially elliptic flow. Thus, the additional length scale,
characteristic of hard collisions, interferes with our ability to extract bulk
matter properties due to the finite volume in space-time subtended by heavy
ion collisions.

By increasing the sampling ratio in Boltzmann treatments while simultaneously
reducing the cross section, one can effectively eliminate these effects in
models. Aside from simplifying the analysis, large sampling ratios are
attractive as they eliminate sensitivities to a variety of arbitrary choices
inherent to simulations and solve a variety of problems related to acausal
propagation. However, in nature particles indeed interact over a finite range,
either by exchange of off-shell particles or through a mutual interaction
through classical fields. Therefore, it is important to understand the degree
to which these effects are physical as opposed to representing numeric
artifacts.

Determining the correct range for non-local interactions in the first fm/c of a
collision might require correctly taking into account both relativity and
quantum effects. For instance, one might determine the extent by considering
the transfer of momentum and energy through exchange of off-shell particles in
the context of a Wigner decomposition of the propagator
\cite{browndanielewicz}. This object might then be analyzed in a matter
analogous to the method in which the classical scattering kernel was studied in
Sec. \ref{sec:collkernel}. However, this is beyond the scope of the current
paper.

On the optimistic side, non-local effects can be interpreted in terms of
viscous parameters as shown in Sec. \ref{subsec:hydro}. This allows the
incorporation of non-local effects into hydrodynamic codes in a
straight-forward manner.  Knowing the viscous parameters would also provide
criteria for tuning Boltzmann algorithms so that they are consistent with
quantum transport considerations.  The tuning could be accomplished by changing
either the sampling factor or by adjusting scattering algorithms. The simple
manner with which the viscous parameters scale with density and cross-section
should simplify such a procedure.

As transport theories address the first one fm/c of a relativistic heavy ion
collision, the role of non-local interactions becomes increasingly
important. For times above 2 fm/c, it is unlikely that the non-localities play
any significant role as the effects scale as $\tau^{-3}$.  Since the nuclei
pass one another at RHIC in less than 0.2 fm/c, whereas cross sections might
approach a square fm, non-local effects might provide a non-negligible source
of stopping as the viscous drag converts longitudinal collective velocity to
heat. The role for such effects in the stopping phase at LHC collisions should
be even greater.

Finally, we mention the effect of non-local interactions one might expect in
the earliest moments of the big bang. In that case cross sections should become
perturbative and particles should be approximately massless. In this case cross
sections should scale as $\alpha^2/T^2$, where $\alpha$ is the unified coupling
constant. Since $T$ would scale as $1/\tau$, the Navier-Stokes viscosity and
the non-local viscosity should both scale identically with $\tau$. Since the
non-local terms scale as $\sigma^2$ while the Navier-Stokes terms scale as
$\sigma^{-1}$, the terms would differ in importance by a constant proportional
to $\alpha^{6}$. Thus, if the system becomes perturbative, one expects the
non-local terms to be negligible compared to the Navier-Stokes terms. However,
the physics of high-energy hadronic collisions is far from perturbative, and
the large coupling constants magnify the importance of the non-local terms
relative to Navier-Stokes terms.

$\tilde{}$

\acknowledgments{ This work was supported by the National Science Foundation,
grant PHY-00-70818, and by the Department of Energy under Contract
No. DE-AC02-98CH10886 and Contract No. DE-FG-02-93ER-10761. B. Zhang's work was
partly supported by the Arkansas Science and Technology Authority Grant
No. 01-B-20.}

\newpage

\end{document}